\documentclass[
aps,
amsmath,amssymb,
prb,
reprint,
groupedaddress,
nolongbibliography
]{revtex4-2}
\usepackage{graphicx}
\usepackage{dcolumn}
\usepackage{bm}
\usepackage{color}

\begin{document}

\title{Single electron routing in a silicon quantum-dot array}

\author{Takeru Utsugi}
\email[]{takeru.utsugi.qb@hitachi.com}
\affiliation{Research \& Development Group, Hitachi, Ltd., Kokubunji, Tokyo 185-8601, Japan
}
\author{Takuma Kuno}
\affiliation{Research \& Development Group, Hitachi, Ltd., Kokubunji, Tokyo 185-8601, Japan
}
\author{Noriyuki Lee}
\affiliation{Research \& Development Group, Hitachi, Ltd., Kokubunji, Tokyo 185-8601, Japan
}
\author{Ryuta Tsuchiya}
\affiliation{Research \& Development Group, Hitachi, Ltd., Kokubunji, Tokyo 185-8601, Japan
}
\author{Toshiyuki Mine}
\affiliation{Research \& Development Group, Hitachi, Ltd., Kokubunji, Tokyo 185-8601, Japan
}
\author{Digh Hisamoto}
\affiliation{Research \& Development Group, Hitachi, Ltd., Kokubunji, Tokyo 185-8601, Japan
}
\author{Shinichi Saito}
\affiliation{Research \& Development Group, Hitachi, Ltd., Kokubunji, Tokyo 185-8601, Japan
}
\author{Hiroyuki Mizuno}
\affiliation{Research \& Development Group, Hitachi, Ltd., Kokubunji, Tokyo 185-8601, Japan
}
\date{\today}

\begin{abstract}
The ability to transport single electrons on a quantum dot array dramatically increases the freedom in designing quantum computation schemes that can be implemented on solid-state devices. So far, however, routing schemes to precisely control the transport paths of single electrons have yet to be established. Here, we propose a silicon single-electron router that transports pumped electrons along the desired route on the branches of a T-shaped quantum dot array by inputting a synchronous phase-controlled signal to multiple gates. Notably, we show that it is possible to achieve a routing accuracy above 99\% by boosting the accuracy of the electron-transport timing with an assist gate in front of the branching paths. We also evaluated the minimum error rate of routing by the model of electron transport based on the Wigner representation in an energy-time space. The results suggest new possibilities for fast and accurate transport of single electrons on the two-dimensional quantum dot arrays.
\end{abstract}
\maketitle

\section{\label{intro}Introduction}
Toward large-scale quantum computations, field programmable qubit arrays (FPQAs), analogous to the classical field programmable gate array (FPGA)~\cite{Boger2023}, have been realized in the form of trapped ion or neutral-atom qubit arrays~\cite{kielpinski2002architecture,pino2021demonstration,bluvstein2022quantum,barnes2022assembly}. In these quantum processors, qubits are spatially transported with high controllability. 
{Similarly, for our ultimate goal of realizing a quantum computer based on large-scale quantum dot (QD) arrays in solid-state devices~\cite{Loss1998,chatterjee2021semiconductor}, the technology for transporting single electrons to an arbitrary QD is one of the most promising technologies. This is because the possibility of dramatically increasing the freedom in the design of quantum computation schemes by electron transport has recently been proposed.~\cite{taylor2005fault,vandersypen2017interfacing,Boter2022,Jnane2022,seidler2022conveyor,Langrock2023}.}

One of the key components is the single-electron router that programmatically controls the electron transport pathway. So far, there are two approaches to routing electrons in solid-state devices, i.e., passive and active ~\cite{bauerle2018coherent,Edlbauer2022}. Passive approaches, e.g., a quantum point contact~\cite{Bocquillon2012,Bocquillon2013,Dubois2013,jullien2014quantum} or a static potential barrier~\cite{Ubbelohde2015}, have been used in studies on electron quantum optics. Alternatively, tunnel-coupled wires can be used to control the paths of flying electron qubits with surface acoustic waves ~\cite{yamamoto2012electrical,takada2019sound,Ito2021}. Although these passive electron-path control techniques have attracted much attention, programmable routing is a nontrivial issue with these techniques. On the other hand, an active approach that is suitable for programmable single-electron routing is the time-gating technique~\cite{Fletcher2013}. This technique controls the electron route by synchronizing a potential barrier with the single-electron source. In fact, it has led to outstanding achievements in studies on single-electron transport, e.g., by enabling analyzes of single-electron wave packets and phonon emissions by using quantum Hall edge states in devices based on GaAs with strong magnetic fields~\cite{Fletcher2013,Waldie2015,Kataoka2016,Johnson2017,Johnson2018,Fletcher2019}. So far, however, these technologies have not been applied to QD arrays.

Inspired by the time-gating technique~\cite{Fletcher2013,Fletcher2019}, we developed a single-electron router on a {T-shaped} QD array in silicon, where we utilized a tunable barrier single-electron pump (SEP)~\cite{Blumenthal2007,Kaestner2008,fujiwara2008nanoampere,Pekola2013,Kaestner2015,Giblin2019} constructed on the QD array as a single-electron source. In this paper, we first describe a SEP operating at 100 MHz based on the previous study~\cite{Utsugi_2023}. Then, we program a single-electron train periodically emitted from the SEP into the right and left paths in the branches ahead, and the single-electron routing operation was achieved by inputting synchronous signals to multiple gates in the QD array. We note that electrons tend to stagnate in silicon QD arrays, which do not have a driving force for electrons like Hall edge channels have and must be driven for proper transport. We show the importance of an assist gate just before the branching path. Namely, the additional signal applied to the assist gate boosts the accuracy of the electron transport timing and improves the routing signal. 
Next, we describe a demonstration of reliable routing of electrons at rates up to 100 MHz. We also describe a method for evaluating the performance of the single-electron router. Conditions for stable operation and operating voltage margins are discussed, and the conclusion is that the single-electron router is on the verge of practical application. We consider that the single-electron router can be applied not only to electron loading~\cite{Volk2019,Mills2019} or spin shuttling~\cite{Baart2016,Fujita2017,Nakajima2018,Yoneda2021,Noiri2022,zwerver2022shuttling} in large-scale two-dimensional QD arrays~\cite{Flentje2017,Ansaloni2020,gilbert2020single,duan2020remote,Mortemousque2021,borsoi2022shared}, but also to current standards based on SEP that use error detection and correction to improve SEP performance ~\cite{yamahata2011accuracy,Yamahata2014,Wulf2013,Fricke2014,Tanttu2015,giblin2016high,ghee2023fidelity}.

\section{Device and experimental setup}
\begin{figure}
\includegraphics[width=80mm]{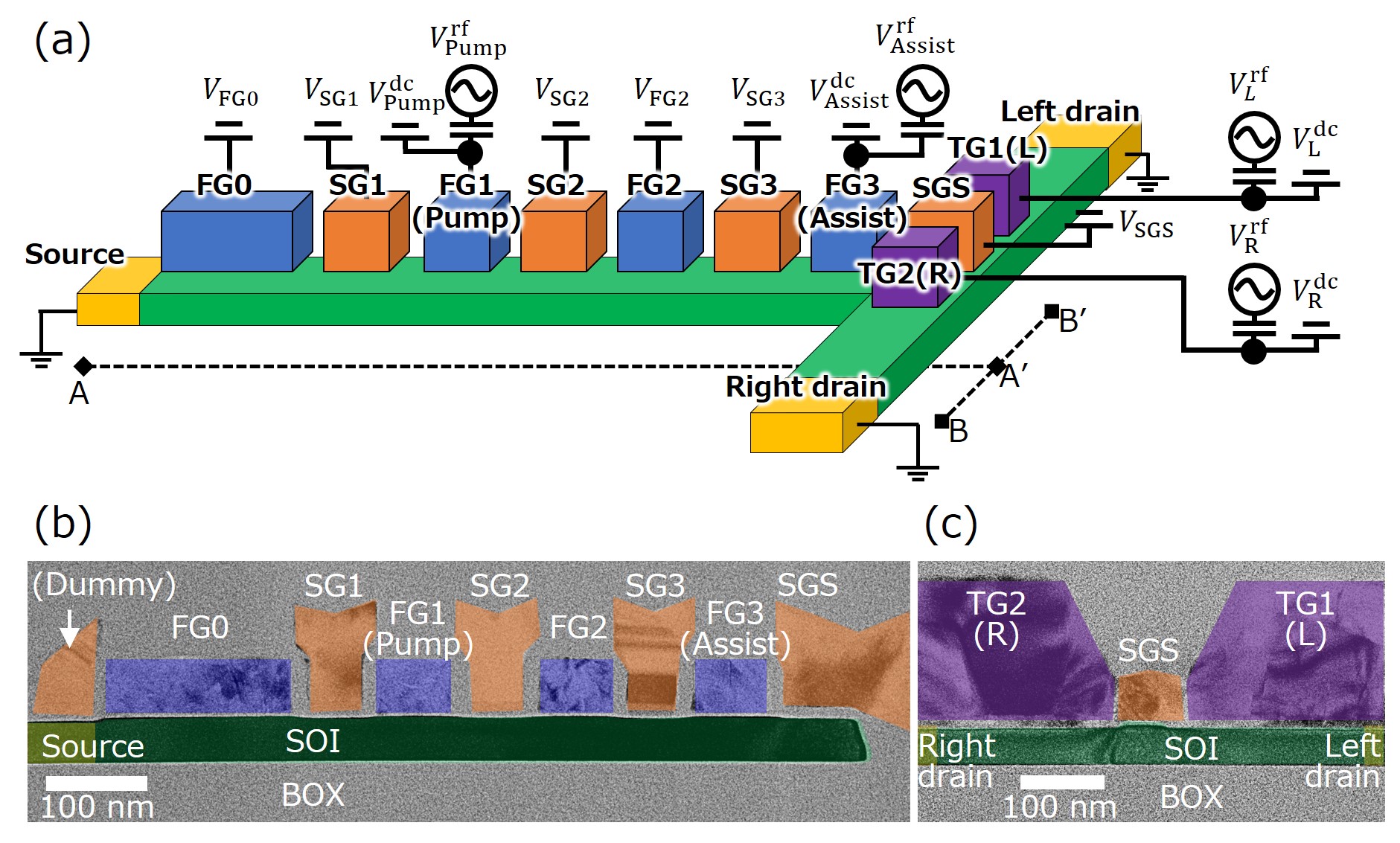}
\caption{\label{0001}Device configuration: (a) Schematic diagrams of the T-shaped device and the measurement setup and TEM images of (b) the A-A’ cross-section and (c) B-B’ cross-section shown in (a).}
\end{figure}

Figure~\ref{0001}(a) shows a schematic diagram of the device configuration used in our study. This device is fabricated with a fully depleted silicon-on-insulator (SOI) wafer and has a T-shaped intrinsic Si channel (green)~\cite{hisamoto2023}. The Si channel has three terminals (yellow): the source and the drains on the right and left sides. Three layers of gate electrodes are formed above the Si channel, e.g., first gates (FG0-FG3, blue), second gates (SG1-SG3 and SGS, orange), and third gates (TG1 and TG2, purple). The second gates are formed by the self-aligned patterning process used to make the first gates, and the third gates are formed by the self-aligned patterning process for the second gates. Figure~\ref{0001}(b) and~\ref{0001}(c) show TEM images in cross-sectional views of the device shown in Fig.~\ref{0001}(a). Table~\ref{004} lists the device parameters~\cite{hisamoto2023}.

\begin{table}[h]
\caption{\label{004}Device parameters.}
\begin{ruledtabular}
\begin{tabular}{lccc}
&\textrm{FG1-3}&\textrm{SG1-3}&\textrm{TG1,2}\\
\colrule
$T_{\rm{ox}}$ & 5 nm & 15 nm & 15 nm\\
$W$ & 50 nm & 50 nm & 50 nm\\
$L$ & 80 nm & 50 nm & 200 nm\\
$T_{\rm{SOI}}$ & 50 nm & 50 nm & 50 nm\\
$T_{\rm{BOX}}$ & 145 nm & 145 nm & 145 nm\\
Gate & B-doped & P-doped & P-doped\\
Channel & \multicolumn{3}{c}{Non-doped Si}\\
\end{tabular}
\end{ruledtabular}
\end{table}

The experimental setup is also schematically shown in Fig~\ref{0001}(a). Note that all measurements were performed at $4$ K {using a dilution refrigerator}. Electrons are sent from the source to the right or left drain, where the bias voltages between the source, right drain, and left drain are set to 0 V in the single-electron routing operation. The currents are measured with a source-measure unit connected to the source, right, and left drains. The voltages applied to the gates are generated by an arbitrary waveform generator. Three gates, FG1, SG2, and FG2, are used as a SEP: inputting an rf signal to FG1 periodically extracts a single electron from the source, whereas SG2 and FG2 are used as adjustment knobs of the SEP operation. {Hereafter, gate FG1 is called the pump gate.} Here, to act as extensions of the source electrode, gates FG0 and SG1 are set to turn-on by applying 2 V. Gates TG1 and TG2 are used as switching gates in a time-gating manner; these are a pair of shutters that select the electron transport path on the branching path of the T-shaped Si channel. The signals are synchronized to the same frequency as the SEP. The phases of the input signals to TG1 and TG2 are inverted so that one is closed when the other is open. {Hereafter, gate TG1 (TG2) is called the L (R) gate.} In addition, gate FG3 is used as an assist gate to assist the routing operation, as described below.  {Hereafter, gate FG3 is called the assist gate.} Gate SG3 is set to appropriate voltages to form the electron pathway. The applied voltage to the gate SGS is adjusted so that no current flows from the right drain to the left drain or vice versa. 

{The rf sinusoidal signals of the same frequency are input to the pump gate (FG1) for the SEP, L gate (TG1) and R gate (TG2) for the switching gates, and the assist gate (FG3).} Single-electron routing operation is achieved by adjusting the rf input phase to each gate, where bias tees are used for applying both dc and rf voltages. These voltages are written as $V_k(t)=V_k^{\rm{dc}}+V_k^{\rm{rf}}\cos(2\pi f t-\phi_k)$ ($k \in \{\rm{Pump, Assist, L, R}\}$), where $V_k^{\rm{dc}}$, $V_k^{\rm{rf}}$, $\phi_k$, $f$, and $t$ are dc voltages, ac amplitudes, initial phase, frequency, and time, respectively. The dc voltages $V_l$ ($l \in \{\rm{FG0, SG1, SG2, FG2, SG3, SGS}\}$) are applied to the other gates. 


\section{Characteristics of single-electron pump, assist gate, and switching gates}
\begin{figure}[h]
\includegraphics[width=80mm]{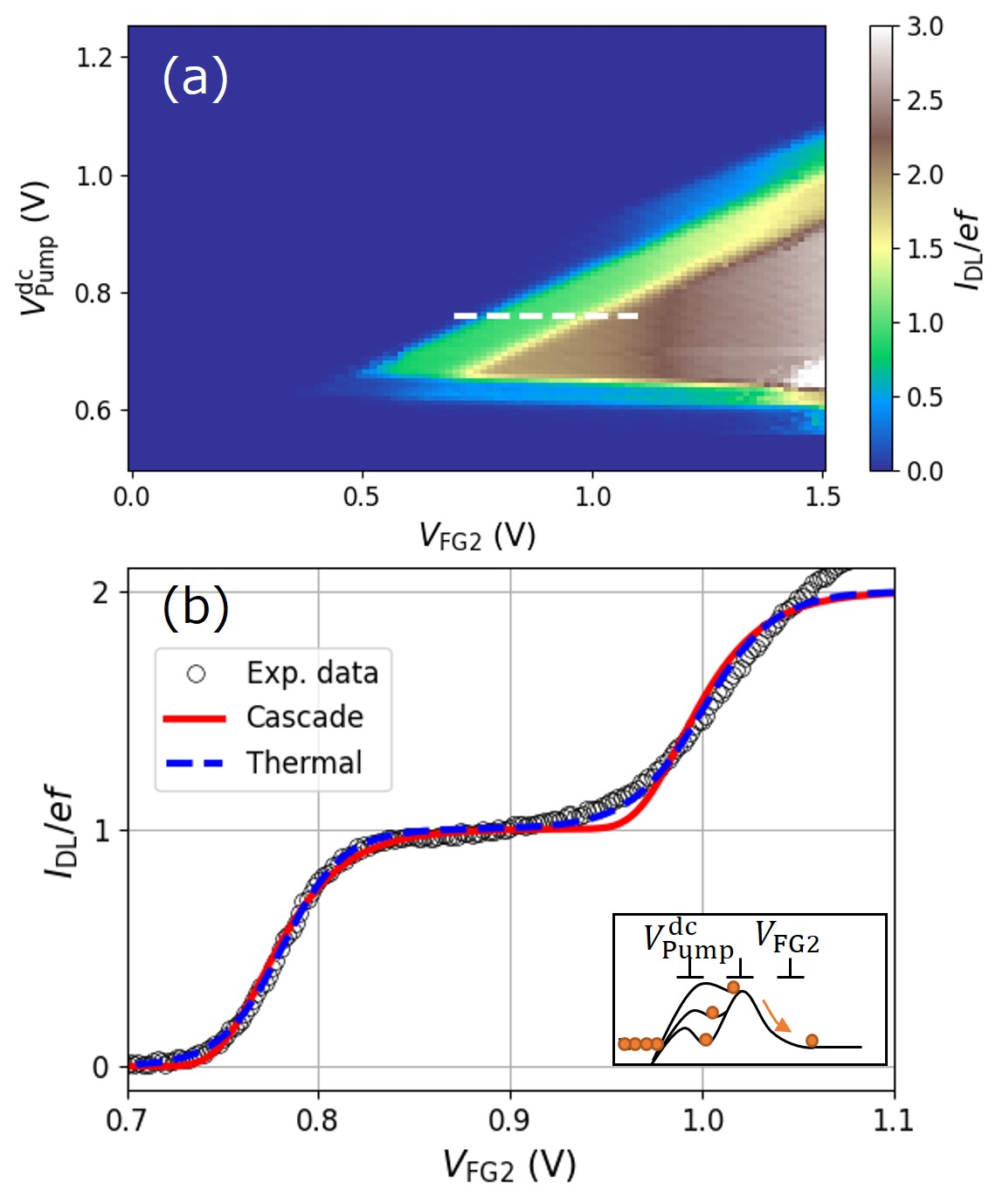}
\caption{\label{0002}Evaluation of the SEP. (a) Current map as a function of dc voltage at the pump gate and FG2. (b) Single-electron plateau observed in the cross-section of the white dashed line in (a), where $V_{\rm{Pump}}^{\rm{dc}}=0.76$ V. {The red solid (blue dashed) line shows a fit by the decay-cascade (thermal-equilibrium) model and the inset shows a schematic diagram of the SEP. The voltage conditions are $V_{\rm{Pump}}^{\rm{rf}}=$ 1.1 V, and $V_{\rm{SG2}}=$ -1.1 V. The right drain is set to open, and the other gates are applied 2 V except for $V_{\rm{FG2}}$.}}
\end{figure}

First, we evaluated the characteristics of the SEP that were described in the previous study~\cite{Utsugi_2023}. Figure~\ref{0002}(a) shows the pumped current as a function of dc voltage at the pump gate and FG2. The electrons flowed from the source to the left drain by setting the right drain to open and the other gates to turn on by 2 V. The left drain current, denoted as $I_{\rm{DL}}$, is normalized by $ef$ in this figure since the pumped current is $ef$, where $e$ is the elementary charge, and $f$ is the operating frequency. We set the operating frequency to $f=100$ MHz in consideration that the cutoff frequency of the input signal to the pump gate is about $250$ MHz. The frequency characteristics of the rf signal input to the pump gate and pumped currents are shown in Appendix~\ref{AppA}. The dc voltage for SG2 and the ac amplitude for the pump gate were set to $V_{\rm{SG2}}=-1.1$ V and $V_{\rm{Pump}}^{\rm{rf}}=1.1$ V, respectively. By setting the dc voltage of the pump gate to $V_{\rm{Pump}}^{\rm{dc}}=0.76$ V (white dashed line), a single electron plateau appeared as shown in Fig.~\ref{0002}(b).

{We fit the experimental data in Fig.~\ref{0002}(b) by two representative models: the decay-cascade model~\cite{fujiwara2008nanoampere,Giblin2012,Kaestner2015,Giblin2019} and thermal-equilibrium model~\cite{Yamahata2014,Zhao2017}.
As shown in Fig.~\ref{0002}(b), the experimental data is more accurately described by the thermal-equilibrium model than the decay-cascade model.
Here, the decay-cascade model is written as}
\begin{equation}
\frac{I_{\rm{DL}}}{ef}=\sum_{i=1,2} \exp \left[ -\exp(-\alpha V_{\rm{FG2}}+\Delta_i) \right],
\label{eq1}
\end{equation}
where the best values of the fitting parameters are $\Delta_{1}=32.5$, $\Delta_{2}=41.7$, and $\alpha=42.2~\rm{V}^{-1}$. 
{On the other hand, the thermal-equilibrium model is written as
\begin{equation}
\frac{I_{\rm{DL}}}{ef}=\sum_{i=1,2} \frac{1}{1+\exp(\beta_i V_{\rm{FG2}}+\gamma_i)},
\label{eq1.1}
\end{equation}
where the best values of the fitting parameters are $\beta_1=-62.8~\rm{V}^{-1}, \gamma_1=49.0, \beta_2=-53.3~\rm{V}^{-1}$, and $\gamma_2=53.3$.
The fitting errors for these two models are evaluated by the root-mean-square error (RMSE) in the range of $V_{\rm{FG2}}=$ 0.8 V to 1.0 V.
As a result, the RMSE for the decay-cascade model is 0.048, and that for the thermal-equilibrium model is 0.024, which suggests this SEP operates in the situation described by the thermal-equilibrium model.
We estimated the minimum error rate with the optimal working point given by the point of inflection of the fit line~\cite{Zhao2017}, resulting in $P_{\rm{error}}^{\rm{Pump}}=4.1\times10^{-5}$, i.e., below 0.01\%, which is good enough for demonstrating the single-electron router.}

\begin{figure}[h]
\includegraphics[width=70mm]{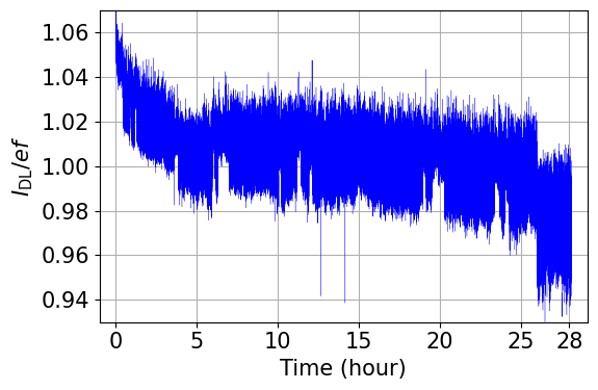}
\caption{\label{A003}Stability evaluation of SEP operating at 100 MHz. The accuracy estimated from the average current $I_{\rm{DL}}$ over 28 hours is 99.86\% (an error of 0.14\% for $ef$), {and the standard deviation of the current over 28 hours is 0.016 (1.6\% for $ef$). The voltage conditions are $V_{\rm{Pump}}^{\rm{rf}}=$ 1.1 V, $V_{\rm{Pump}}^{\rm{dc}}=$ 0.76 V, $V_{\rm{SG2}}=$ -1.1 V, and $V_{\rm{FG2}}=$ 0.88 V. The right drain is set to open, and the other gates are applied 2 V. }}
\end{figure}

{Then we evaluated the stability of the SEP operated at 100 MHz. As shown in Fig.~\ref{A003}, the pumped current, $I_{\rm{DL}}\approx ef$, was nearly constant during a continuous measurement lasting 28 hours. The accuracy estimated from the average value of the current was 99.86\%, indicating an error of 0.14\% for $ef$, and the standard deviation of the current was 0.016 (1.6\% for $ef$). This result is also sufficient for evaluating and demonstrating the single-electron router in this study.
}

\begin{figure}[h]
\includegraphics[width=80mm]{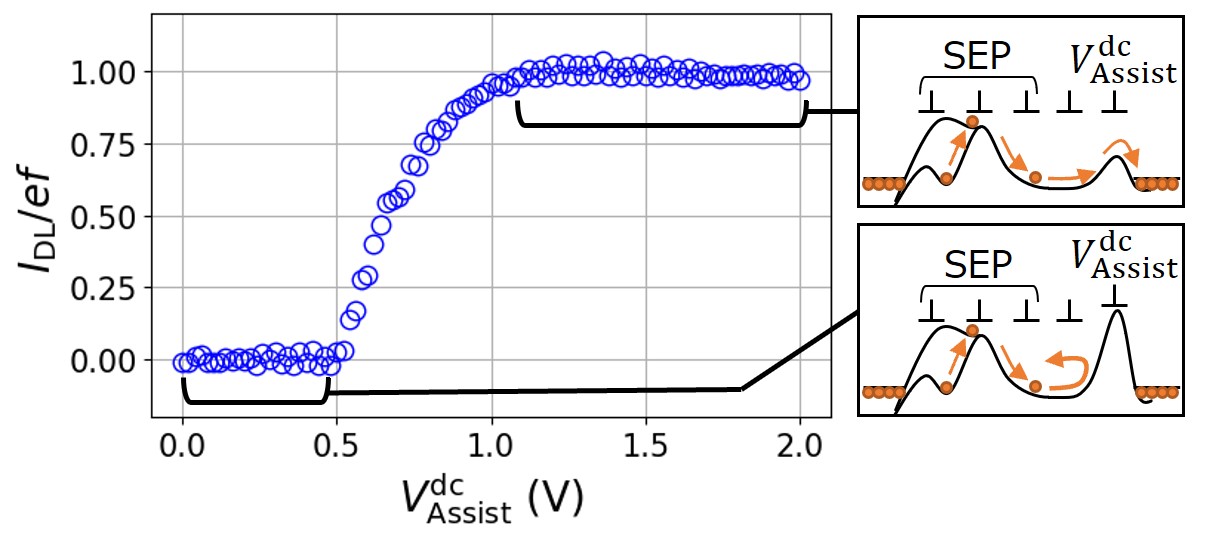}
\caption{\label{0003}The dc characteristics of the assist gate (FG3) for a 100-MHz SEP and the schematic diagram of (upper) blocking and (lower) transmission of electrons. {The voltage conditions are $V_{\rm{Pump}}^{\rm{rf}}=$ 1.1 V, $V_{\rm{Pump}}^{\rm{dc}}=$ 0.76 V, $V_{\rm{SG2}}=$ -1.1 V, and $V_{\rm{FG2}}=$ 0.88 V. The right drain is set to open, and the other gates are applied 2 V except for $V_{\rm{Assist}}^{\rm{dc}}$.}}
\end{figure}

Next, we evaluated the characteristics of the assist gate, which assists in routing the electrons. We drove the SEP at 100 MHz, as indicated above, and measured the dependence of the pumped current on the dc voltage applied to the assist gate. The results are shown in Fig.~\ref{0003}. It was found that the pumped electrons are blocked when the applied voltage is approximately 0.5 V or lower and that electrons are transmitted when the applied voltage is higher than 1.0 V. The presence of blocking electrons indicates that the pumped electrons flow backward after pumping (see the diagram in Fig.~\ref{0003}). 
{This result shows that the applied voltage to the assist gate needed to switch between transmission and blocking is about $0.5$ V difference. In addition, a dc voltage of approximately {1.2 V} or more should be applied to avoid the backward flow of the pumped current.}

\begin{figure}[h]
\includegraphics[width=70mm]{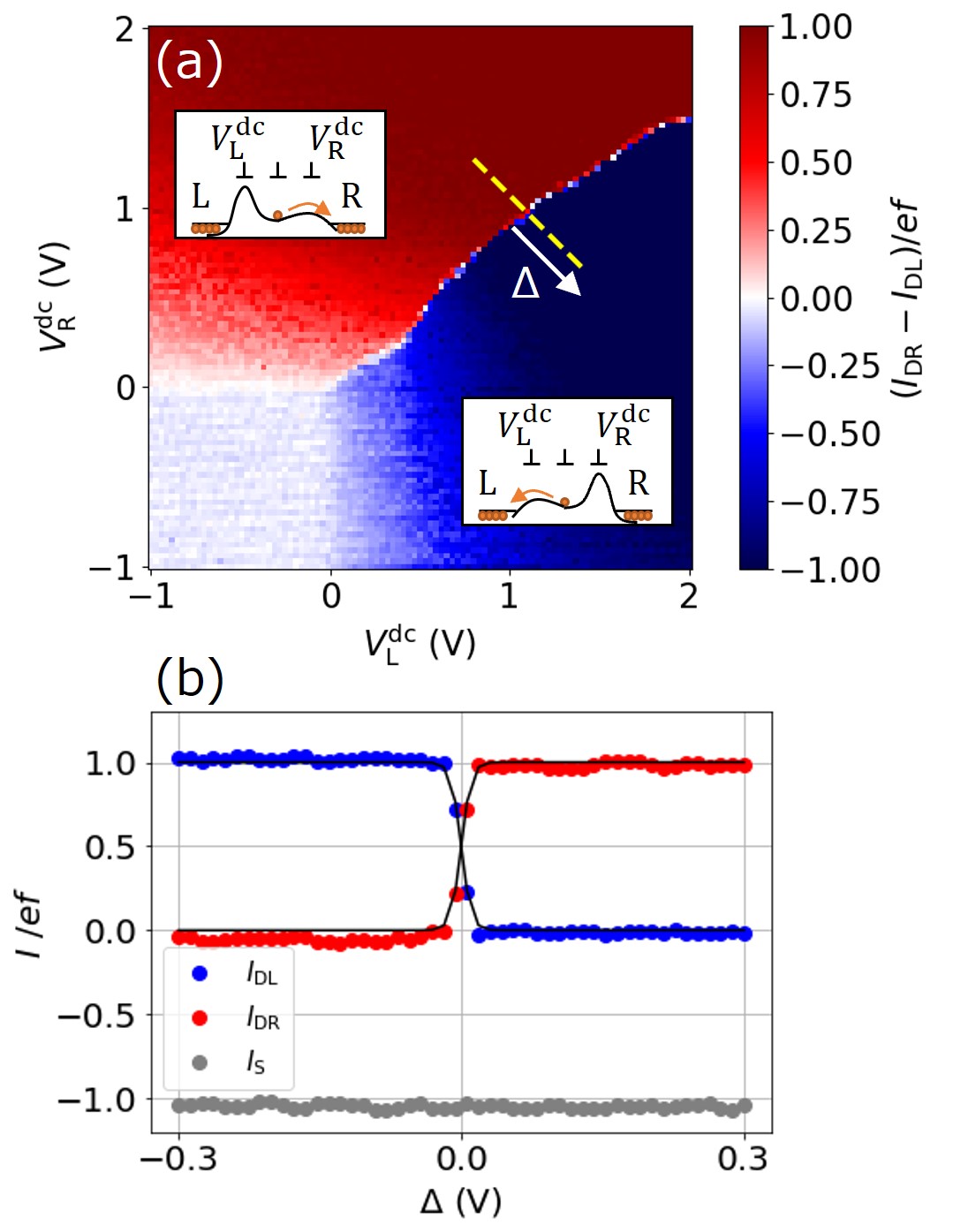}
\caption{\label{0004}The dc characteristics of the switching gates for a 100-MHz SEP. (a) Contour map of the difference between pumped currents flowing from the SEP to the left drain and right drain $(I_{\rm{DR}}- I_{\rm{DL}})/ef$ as a function of the voltages $V_{\rm{L}}^{\rm{dc}}$ and $V_{\rm{R}}^{\rm{dc}}$. The insets are schematic diagrams of switching the electron routes. (b) Currents $I_{\rm{DL}}$, $I_{\rm{DR}}$, and $I_{\rm{S}}$ as a function of the potential detuning $\Delta$ shown by the yellow dashed line in (a), where $V_{\rm{L}}^{\rm{dc}}=1.1-\Delta$ (V) and $V_{\rm{R}}^{\rm{dc}}=0.97+\Delta$ (V). The black lines are fits using the non-ideal step function shown in Eq.~(\ref{eq2}) with $\sigma = -5.2$ mV for $I_{\rm{DL}}$ and $\sigma = +5.2$ mV for $I_{\rm{DR}}$. {The voltage conditions are $V_{\rm{Pump}}^{\rm{rf}}=$ 1.1 V, $V_{\rm{Pump}}^{\rm{dc}}=$ 0.76 V, $V_{\rm{SG2}}=$ -1.1 V, and $V_{\rm{FG2}}=$ 0.88 V. The other gates are applied 2 V except for $V_{\rm{L}}^{\rm{dc}}$ and $V_{\rm{R}}^{\rm{dc}}$.}}
\end{figure}

Then, we evaluated the characteristics of the switching gates. Figure~\ref{0004}(a) is a contour map of the difference between pumped currents flowing from the SEP to the left drain $I_{\rm{DL}}$ and right drain $I_{\rm{DR}}$ as a function of the voltages $V_{\rm{L}}^{\rm{dc}}$ and $V_{\rm{R}}^{\rm{dc}}$ applied to the L and R gates, respectively. The voltages of all the gates except for the SEP and the switching gates were set to 2 V, i.e., to turn them on. In the region where $V_{\rm{L}}^{\rm{dc}}$ ($V_{\rm{R}}^{\rm{dc}}$) is lower than about 0 V, the electron energy at the switching gates is lower than the potential barriers of the switching gates, resulting in that the pumped electrons flow backward after pumping. On the other hand, when $V_{\rm{L}}^{\rm{dc}}$ ($V_{\rm{R}}^{\rm{dc}}$) exceeds about 0 V, the dependence of the current on the left or right drain changes abruptly, indicating that electrons flow toward the lower potential in an almost deterministic manner [see the insets in Fig.~\ref{0004}(a)]. Therefore, it is possible to control which direction electrons are sent by adjusting the heights of the potentials by the switching gates.

Fig.~\ref{0004}(b) shows cross-sections along the potential detuning $\Delta$ between the L and R gates, which is shown as the yellow dashed line in Fig.~\ref{0004}(a). In the region around this line, the electron energy is higher than the potential barriers of the switching gates. Considering the broadening $\sigma$ due to gate voltage noise or energy broadening of electrons, we decided to fit the experimental data as a function of $\Delta$ in Fig.~\ref{0004}(b) with a non-ideal step (Fermi-Dirac-like) function ~\cite{Waldie2015}:
\begin{equation}
T(\Delta)=\frac{1}{\exp(-\Delta/\sigma)+1}.
\label{eq2}
\end{equation}
These characteristics of the electron transport are consistent with typical transmission probability of the single-electron partitioning measurements in two transport channels~\cite{Fletcher2013,Ubbelohde2015,takada2019sound}. Since the experimental data are well fitted by $T(\Delta)$, we took $T(\Delta)$ to be the transmission probability of the electrons.
{Suppose the energy broadening is converted to the electron temperature $T_e$, i.e., $\sigma=k_{\rm{B}} T_e/a$, where $k_{\rm{B}}$ and $a$ are the Boltzmann constant and the lever arm of the switching gates, respectively. From the estimation of $a$ shown in Appendix \ref{AppP}, the fit result of $|\sigma|=5.2$ mV suggests $T_e \sim 10-20$~K, which is reasonable because of the temperature increase due to rf-induced heating added to the base temperature of 4K~\cite{Zhao2017}.}

\section{Operating condition}
\begin{figure}[h]
\includegraphics[width=80mm]{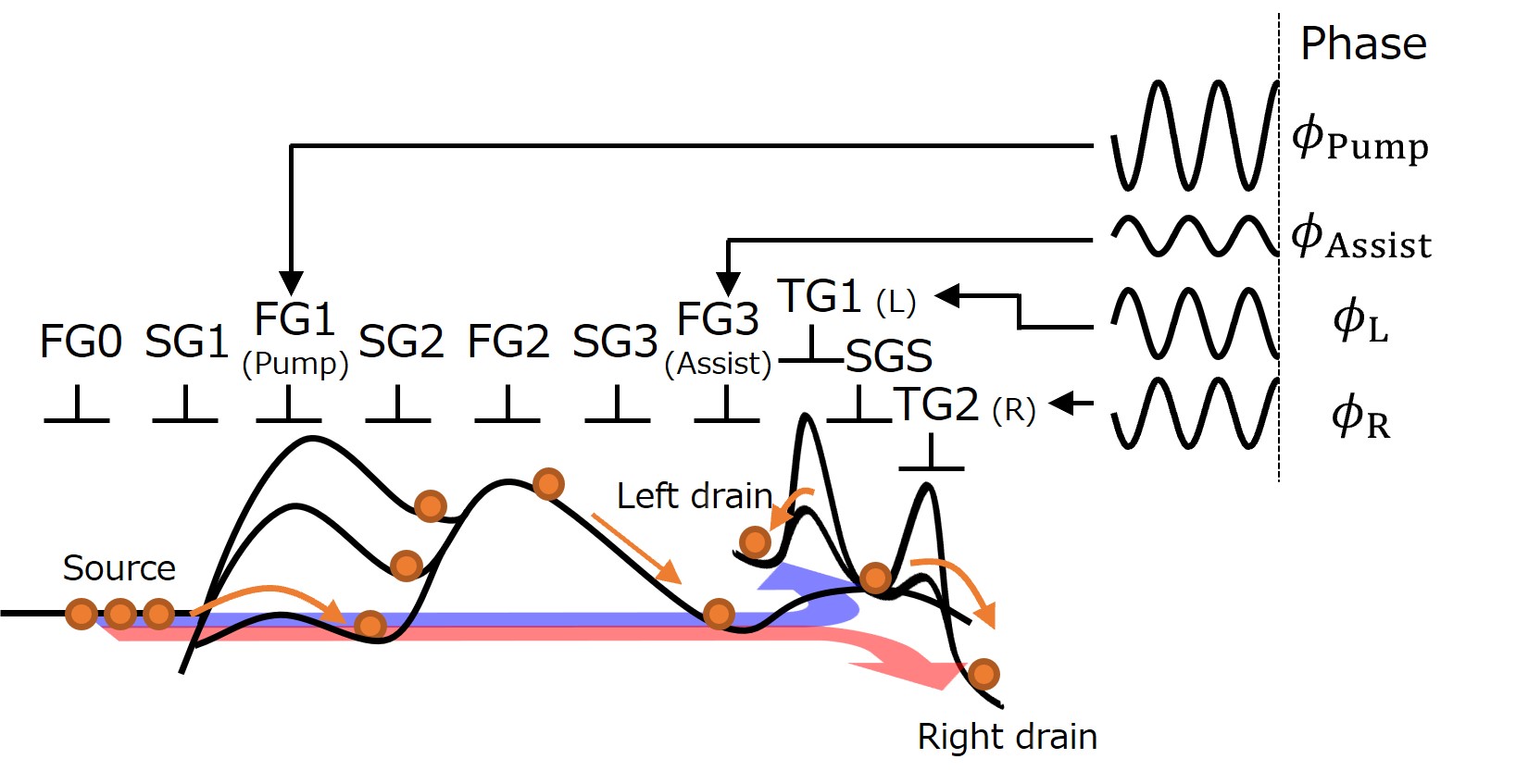}
\caption{\label{0005} Schematic diagram of potentials in the proposed single-electron routing scheme. The rf sinusoidal signals of 100 MHz were applied to the pump, L, R, and assist gate. {The voltage conditions are $V_{\rm{Pump}}^{\rm{rf}}=$ 1.1 V, $V_{\rm{Pump}}^{\rm{dc}}=$ 0.76 V, $V_{\rm{SG2}}=$ -1.1 V, $V_{\rm{FG2}}=$ 0.88 V, $V_{\rm{SG3}}=$ 0.5 V, $V_{\rm{Assist}}^{\rm{rf}}=$ 0 V or 0.1 V, $V_{\rm{Assist}}^{\rm{dc}}=$ 2 V, $V_{\rm{L}}^{\rm{rf}}=V_{\rm{R}}^{\rm{rf}}$ = 0.5 V,  and $V_{\rm{SGS}}=V_{\rm{L}}^{\rm{dc}}=V_{\rm{R}}^{\rm{dc}}$ = 0 V. The phases of the L and R gates were inverted i.e., $\phi_{\rm{R}}=\phi_{\rm{L}}+\pi$ (rad).}}
\end{figure}

\begin{figure}[h]
\includegraphics[width=70mm]{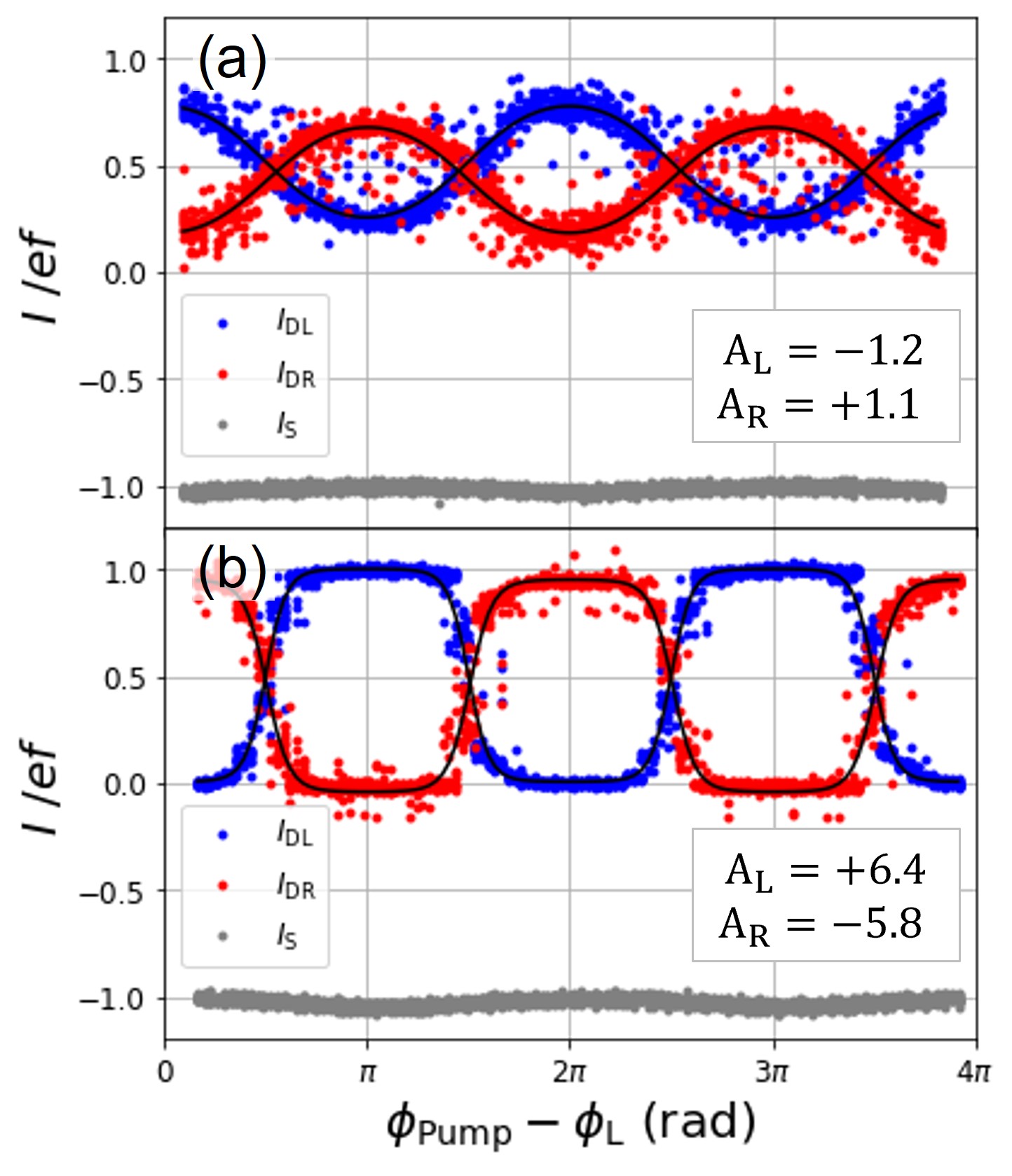}
\caption{\label{0006} Single-electron routing characteristics (a) unassisted and (b) assisted by the rf signal input to the assist gate. The left drain current $I_{\rm{DL}}$ (blue), the right drain current $I_{\rm{DR}}$ (red), and the source current $I_{\rm{S}}$ (gray) are plotted as functions of the phase difference between the SEP and switching gates $\phi_{\rm{Pump}}-\phi_{\rm{L}}$, where the phase of the switching gates is set to $\phi_{\rm{R}}=\phi_{\rm{L}}+\pi$ (rad). In (b), the amplitude is set to $V_{\rm{Assist}}^{\rm{rf}}=0.1$ V and the phase of the input signal applied to the assist gate is in-phase with the SEP, i.e., $\phi_{\rm{Assist}}=\phi_{\rm{Pump}}$. The black lines and insets show a fit and the best-fit result of $A$ by the model curve on the final line of Eq.~(\ref{eq4}), corrected for offset. {The voltage conditions are the same as that of Fig.~\ref{0005}.}}
\end{figure}

The electron flow of the proposed routing scheme is shown in Fig.~\ref{0005}. We controlled the phases of the rf sinusoidal signals input to gates the pump, assist, L, and R gates, i.e., $\phi_{\rm{Pump}}$, $\phi_{\rm{Assist}}$, $\phi_{\rm{L}}$, and $\phi_{\rm{R}}$, respectively. $\phi_{\rm{L}}$ and $\phi_{\rm{R}}$ were inverted, i.e., $\phi_{\rm{R}}=\phi_{\rm{L}}+\pi$ (rad), so that the switching gates would operate as a switch of routes. We set dc voltages to $V_{\rm{SG3}}=0.5$ V and $V_{\rm{Assist}}^{\rm{dc}}=2$ V. $V_{\rm{SGS}}$, $V_{\rm{L}}^{\rm{dc}}$ and $V_{\rm{R}}^{\rm{dc}}$ were set to around 0 V; these voltages were sensitive to experimental disturbances and were adjusted by a few tens of mV to ensure proper routing for each measurement. We also set the ac amplitudes to $V_{\rm{L}}^{\rm{rf}}=V_{\rm{R}}^{\rm{rf}}=0.5$ V. {Note that we have performed all routing experiments by first tuning the voltage condition for SEP alone and then driving the assist and switching gates. Since the SEP is sensitive to ambient voltage conditions, changes in the assist and switching gate voltages cause the pumped current to shift a few percent from $ef$, as seen especially in Figs.~\ref{0006}, \ref{0009}, and \ref{0010}.}

The single-electron routing characteristics without and with rf signal input to the assist gate (unassisted and assisted cases) are compared in Figs.~\ref{0006}(a) and~\ref{0006}(b). In the assisted case, the amplitude and the phase of the input rf signal were set to 0.1 V and 0 radians (in-phase) relative to the SEP. Since the sum of $I_{\rm{DL}}$ and $I_{\rm{DR}}$ equals the source current $-I_{\rm{S}}$ in both cases, the electrons flow to either the right drain or the left drain without a backward flow. The currents $I_{\rm{DL}}$ and $I_{\rm{DR}}$ vary with the phase difference between the SEP and the switching gate in both cases. This fact suggests that the route of the pumped electrons is selected to be either the right drain or the left drain by controlling the phase of the switching gate synchronized with the SEP. However, in the unassisted case shown in Fig.~\ref{0006}(a), the modulation of the routing is relatively low. That is, the peaks of $I_{\rm{DL}}$ and $I_{\rm{DR}}$ are low, indicating that deterministic routing is not achieved. In contrast, the modulation of the routing is more significant in the assisted case shown in Fig.~\ref{0006}(b). Therefore, deterministic routing is achieved by the rf signal input to the assist gate. Note that the phases of the current curves $I_{\rm{DR}}$ or $I_{\rm{DL}}$ in the unassisted and assisted cases are inverted, the reason for which is discussed below.

Now let us examine the current characteristics in the single-electron router based on the model of electron transport in the time-gating technique~\cite{Fletcher2019}. The basic model of the single-electron router is described as
\begin{equation}
\frac{I_{\rm{DL}}}{ef} = \iint W(E,t)T(E,t)dEdt,
\label{eq3}
\end{equation}
where $W(E,t)$ is the Wigner quasiprobability function in the energy-time ($E$-$t$) space of the electron arriving at the switching gates, and $T(E,t)$ is the energy- and time-dependent transmission probability of the barriers controlled by the switching gates. Our model assumes no energy broadening of electrons, i.e., $W(E,t)=\delta(E-E_{\rm{p}})P(t)$, where $E_{\rm{p}}$ is the energy of an electron at the switching gates, and $P(t)$ is the distribution of the electron emission times at the switching gates, the so-called jitter of electron transport. Furthermore, we assume that $P(t)$ is a Gaussian distribution, $P(t)=\frac{1}{\sqrt{2\pi} \xi} \exp \left[-\frac{(t-t_0)^2}{2\xi^2}\right]$, where $t_0$ and $\xi$ are the average and standard deviation of the electron's arrival time, respectively. The transmission probability $T(E,t)$ is basically given by Eq.~(\ref{eq2}). Here, we assume that the electron energy is sufficiently higher than the potential barriers of the switching gates. Accordingly, the detuning $\Delta$ in Eq. (\ref{eq2}) can be written as a time-dependent but electron-energy-independent function, $\Delta(t) = V_0 \cos(2\pi ft - \phi_{\rm{L}})$, where $V_0$ is a constant value reflecting the difference between the input amplitudes of the switching gates. From these assumptions, considering one period $\tau=1/f$ and $t_0=\tau \phi/(2\pi)$ where $\phi$ is the phase reflecting the timing of the electron transport to the switching gate, we can rewrite Eq.~(\ref{eq3}) as
\begin{equation}
\begin{split}
\frac{I_{\rm{DL}}}{ef}
&= \iint W(E,t)T(E,t)dEdt\\
&\approx \iint \delta(E-E_p) P(t) T(E,t)dEdt\\
&= \int_0^\tau P(t) T(E_{\rm{p}},t)dt\\
&=\frac{1}{\sqrt{2\pi} \xi} \int^{\tau}_{0} \frac{\exp \left[-\frac{\left(t-\frac{\tau \phi}{2\pi}\right)^2}{2\xi^2}\right]}{\exp \left[-\frac{V_0}{\sigma} \cos(2\pi ft - \phi_{\rm{L}}) \right]+1} dt \\
&\approx \frac{1}{\exp \left[A \cos(\phi-\phi_{\rm{L}})\right]+1},
\end{split}
\label{eq4}
\end{equation}
where $A$ is a constant value that reflects the single-electron router's performance, involving the router's signal-to-noise ratio and the jitter of electron transport at the switching gates. The approximation on the fifth line in Eq.~(\ref{eq4}) holds in our experimental situation (this is shown in Appendix \ref{AppD}). On the fourth line in Eq.~(\ref{eq4}), $|V_0/\sigma|$ indicates the signal-to-noise ratio of the single-electron router, since $V_0$ is proportional to the input voltage amplitude and $\sigma$ is proportional to the voltage noise of switching gates (we assumed no energy broadening of electrons). For example, in the no jitter case, i.e., $\xi \rightarrow 0$, $P(t)$ becomes a delta function, $P(t)=\delta(t-t_0)$. In this case, $A=-V_0/\sigma$ is straightforwardly derived. In addition, by flipping the sign of $A$, we can derive a similar form for $I_{\rm{DR}}$. The experimental data on $I_{\rm{DL}}$ and $I_{\rm{DR}}$ as a function of the phase difference $\phi-\phi_{\rm{L}}$ can be fitted by the final approximation in Eq.~(\ref{eq4}) with the fitting parameter $A$. Therefore, we can evaluate the routing performance by using the fitting results of $A$.
{Note that we confirmed by simulation that $A$ is valid for roughly less than 50 since the phase modulation steps are finite in our experiment (see Appendix~\ref{AppX}).}

The absolute value of the parameter, $|A|$, directly reflects the minimum error rate. When $\phi_{\rm{Pump}}-\phi_{\rm{L}}=\pi$ and $A>0$, $I_{\rm{DL}}$ takes the maximum value of $ef/(1+e^{-A})$. Accordingly, the minimum error rate of the single-electron router $P_{\rm{error}}^{\rm{Router}}$ can be evaluated as $P_{\rm{error}}^{\rm{Router}}=1/(1+e^{|A|})$. Namely, a larger value of $|A|$ lowers the routing error; for example, $|A|>4.6$ ($|A|>6.9$) is required to achieve an error rate of less than 1\% (0.1\%).

Best fit results are also shown in Fig.~\ref{0006}, where the current curves are fitted by the final approximation in Eq.~(\ref{eq4}). Here, we set $\phi_{\rm{Pump}}$ into $\phi$  and corrected for the offset. The parameter $A$ is obtained for $I_{\rm{DL}}$ ($I_{\rm{DR}}$) as $A_{\rm{L}}$ ($A_{\rm{R}}$). The result that the parameters $|A_{\rm{L}}|$ and $|A_{\rm{R}}|$ in the assisted case are larger than in the unassisted case confirms that the error rate of routing can be reduced by the rf signal input to the assist gate. Furthermore, the results in the assisted case are $|A_{\rm{L}}|=6.4$ and $|A_{\rm{L}}|=5.8$, which are larger than $4.6$, so it is possible to achieve a routing accuracy of above 99\% by implementing assist gates in front of the branching paths. 
{Note that $I_{\rm{DR}}$ in Fig.~\ref{0006}(b) is fitted using the model curve with an offset term added, and the fit curve deviates a few percent from $ef$ around $2\pi$ on the horizontal axis. Nevertheless, the routing performance is evaluated by $|A|$ because the deviation from $ef$ results from the performance degradation of the SEP due to the ambient voltage condition, not that of the routing.}

\begin{figure}[h]
\includegraphics[width=75mm]{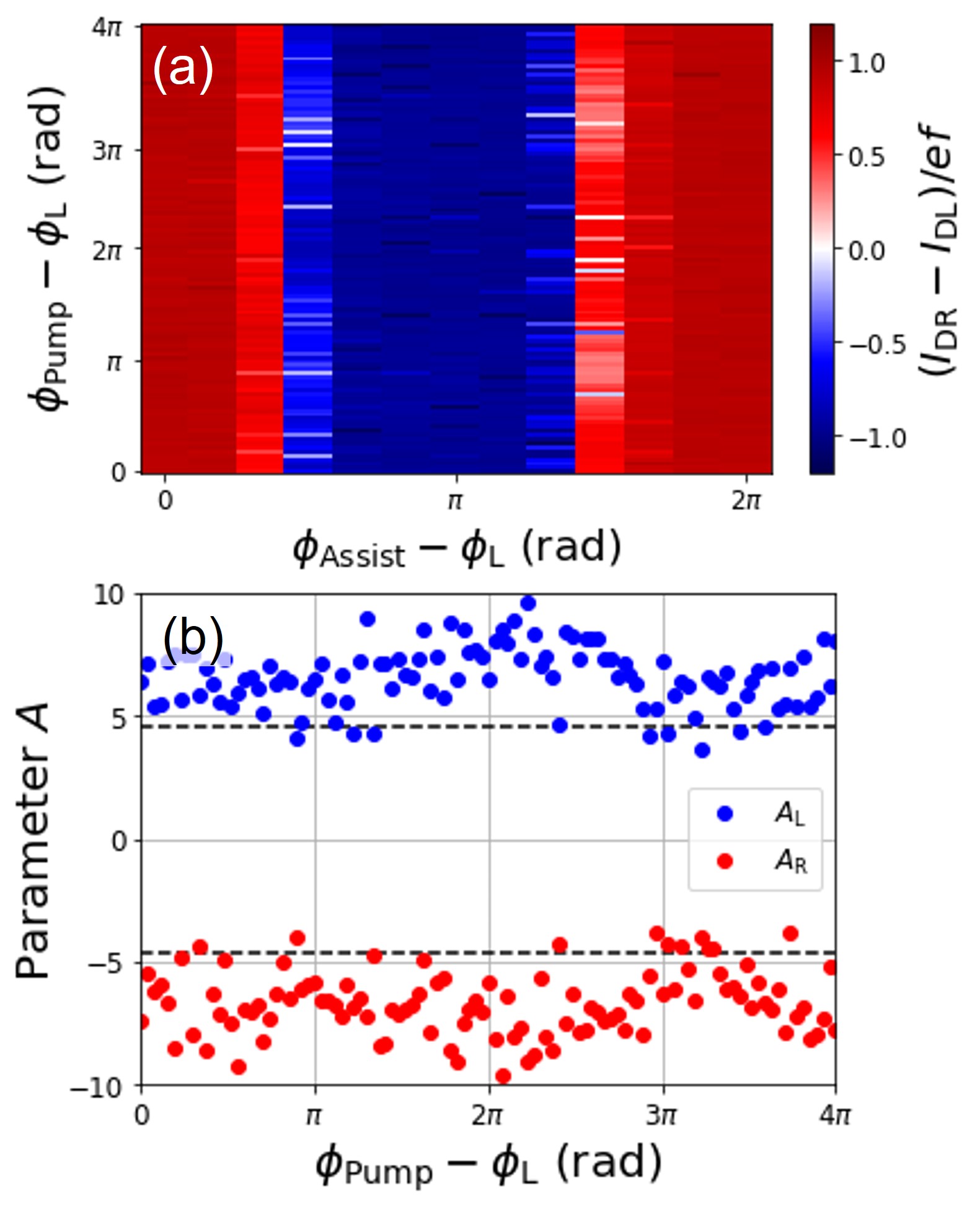}
\caption{\label{0007}Phase dependence of single-electron routing: (a) Contour map of the difference between pumped currents flowing from the SEP to the left drain and the right drain $(I_{\rm{DR}}- I_{\rm{DL}})/ef$ as a function of the phase difference between the SEP and the assist gate, where $V_{\rm{Assist}}^{\rm{rf}}=0.1$ V. (b) The parameter $A$ obtained by fitting the current curves in the horizontal cross-section in (a). This result indicates that most of the parameters $|A_{\rm{L}}|$ and $|A_{\rm{R}}|$ are larger than 4.6 (black dashed line), representing $P_{\rm{error}}^{\rm{Router}}$=1\%; i.e., the routing depends on the phase of the assist gate. {The voltage conditions are the same as that of Fig.~\ref{0005}.}}
\end{figure}

To determine which phase of the SEP or the assist gate contributes to the routing, Fig.~\ref{0007}(a) shows the contour map of the current difference referenced to $\phi_{\rm{L}}$ and Fig.~\ref{0007}(b) shows the parameter $A$ obtained by fitting the current curves in the horizontal cross-section in Fig.~\ref{0007}(a), where {we set $\phi_{\rm{Assist}}$ into $\phi$ in Eq.~(\ref{eq4})}. Here, the amplitude of the rf signal applied to the assist gate was set to 0.1 V. These results show that the routing does not depend on $\phi_{\rm{Pump}}$, but is determined only by $\phi_{\rm{Assist}}$ when an rf signal is input to the assist gate. This fact suggests that the pumped electrons are stored once they come near the assist gate and are pushed out again by the assist gate operation; at that moment, they are sent to the route on the low-potential side. On the other hand, when there is no rf signal input to the assist gate, the electrons are pushed out and routed when the pump injects the next electron because of the Coulomb interaction of the electrons; thus, we can interpret that the phase of the current curves in Fig.~\ref{0006}(a) is inverted.

\begin{figure}[h]
\includegraphics[width=80mm]{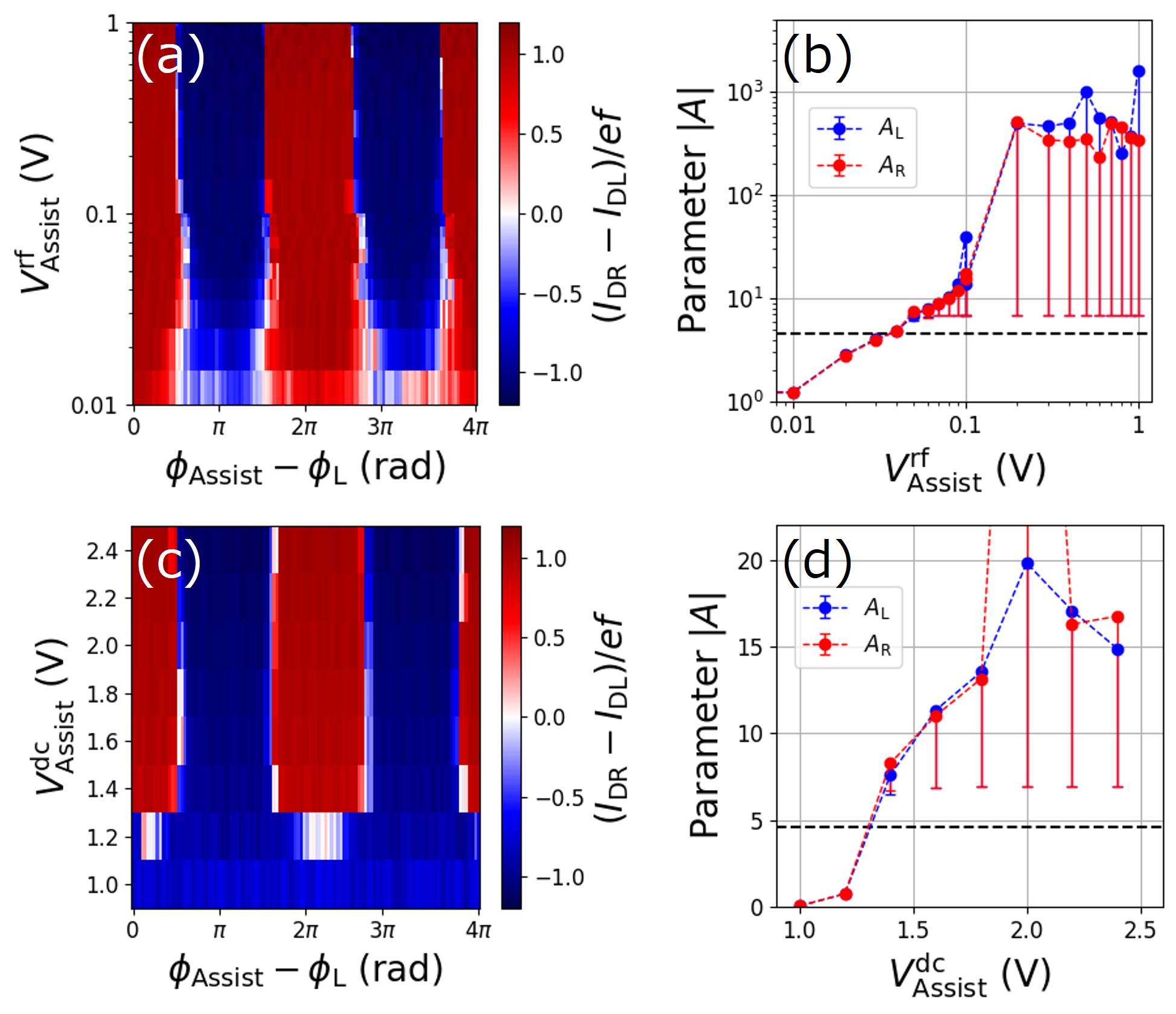}
\caption{\label{0008} (a) Current map as a function of the phase difference $\phi_{\rm{Assist}}-\phi_{\rm{L}}$ and the rf amplitude $V_{\rm{Assist}}^{\rm{rf}}$, where $V_{\rm{Assist}}^{\rm{dc}}=2$ V. (b) Parameter $|A|$ as a function of $V_{\rm{Assist}}^{\rm{rf}}$, estimated by fitting the horizontal cross-section shown in (a). (c) Current map as a function of the phase difference $\phi_{\rm{Assist}}-\phi_{\rm{L}}$ and the dc voltage applied to the assist gate $V_{\rm{Assist}}^{\rm{dc}}$, where $V_{\rm{Assist}}^{\rm{rf}} = 0.1$ V. (d) Parameter $|A|$ as a function of $V_{\rm{Assist}}^{\rm{dc}}$, estimated by fitting the horizontal cross-section shown in (c). In (b) and (d), the black dashed lines show $A=4.6$ representing $P_{\rm{error}}^{\rm{Router}}$=1\%{, and error bars indicate the range of the parameter $|A|$ where $P_{\rm{error}}^{\rm{Router}}$ degrades by 0.1\% with $|A|-\ln\{[{1}/{(e^{|A|}+1)}+10^{-3}]^{-1}-1\}$. {Parameter $A$ is valid for roughly less than 50 because of the finite phase modulation step (see Appendix~\ref{AppX}).} The voltage conditions are the same as that of Fig.~\ref{0005} except for $V_{\rm{Assist}}^{\rm{rf}}$ and $V_{\rm{Assist}}^{\rm{dc}}$.  The phase conditions are $\phi_{\rm{Pump}}=\phi_{\rm{L}}=\phi_{\rm{R}}+\pi$, and $\phi_{\rm{Assist}}$ is varied.}}
\end{figure}

In addition, let us investigate the operating conditions of the assist gates in detail. The routing performance is plotted as functions of the rf amplitude and dc voltage applied to the assist gate in Figs.~\ref{0008}(a,b) and \ref{0008}(c,d). The amplitude dependence shown in Figs.~\ref{0008}(a,b) indicates that deterministic routing was achieved at an input amplitude of approximately 0.05 V or higher. This result indicates that a potential change of approximately 0.05 V is required for electrons stored near the assist gate to be transported to the right or left drain. 
{This result also indicates that the ability to apply the dc voltage difference necessary for switching between electron transmission and blocking to the assist gate (approximately 0.5 V in this case from Fig. 3) was sufficient for the assist gate to work well in the routing operation.}

The dc voltage dependence shown in Fig.~\ref{0008}(c,d) indicates that deterministic routing was achieved at approximately 1.4 V or higher. Thus, at lower voltages, it is difficult for electrons to be stored near the assist gate, and as a result, routing by pushing out electrons by changing the voltage of the assist gate does not work. 
{The reason for the significant performance degradation below 1.2V is due to the backward flow of the pumped current shown in Fig.~\ref{0003}.}
Consequently, by adjusting the operating voltage of the assist gate appropriately, it was found that routing could be controlled by synchronizing the timing of electron transport with the switching gate after the electrons are stored near the assist gate. Hereafter, we set the assist gate voltage to $V_{\rm{Assist}}^{\rm{dc}}=2$ V and $V_{\rm{Assist}}^{\rm{rf}}=0.1$ V. 
{Note that we aimed to control the voltage amplitude as small as possible within the range where sufficient performance could be obtained. This is a trivial matter in this proof-of-concept experiment, but an indispensable viewpoint for achieving large-scale integration with low power consumption in the future.}

\section{Demonstration of single-electron router}
\begin{figure*}
\includegraphics[width=150mm]{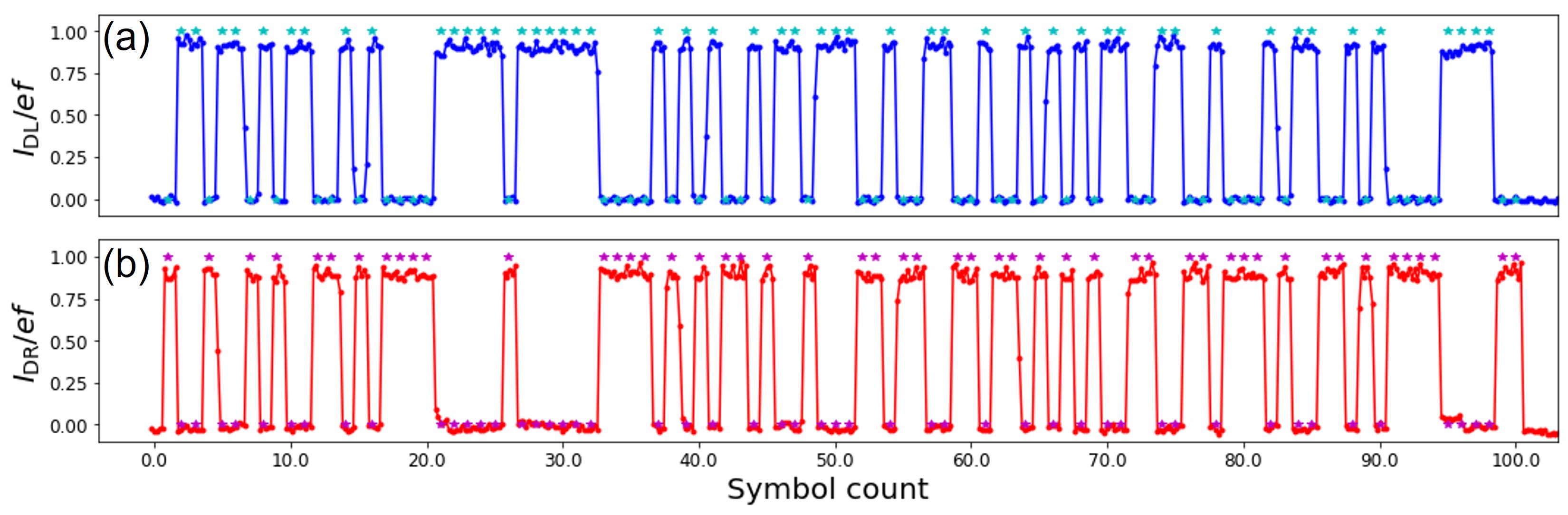}
\caption{\label{0009}Current in (a) the left drain and (b) the right drain of the single-electron router with a 100-MHz SEP. The stars ($\star$) show the programmed symbol. All 100 symbols were accurately routed. {The voltage conditions are the same as that of Fig.~\ref{0005}. The phase conditions are $\phi_{\rm{Pump}}=\phi_{\rm{L}}=\phi_{\rm{R}}+\pi$, and $\phi_{\rm{Assist}}$ is varied.}}
\end{figure*}

Figure~\ref{0009} shows the results of a demonstration of a programmable single-electron routing operation under the conditions obtained above. The SEP was operated at 100 MHz, and the periodically sent electrons were routed to either the right or left drain. As shown in Appendix~\ref{AppC}, a symbol sequence of 100 symbols of binary pseudo-random numbers was generated for R and L, and $\phi_{\rm{Assist}}$ with respect to $\phi_{\rm{L}}$ was set to $\pi$ for `L' ($0$ for `R') according to this symbol sequence. The stars ($\star$) in Fig.~\ref{0009} represent the programmed symbols; they show that accurate routing was achieved for all 100 symbols. Note that the measurement duration per symbol was set to 2 seconds.

\begin{figure*}
\includegraphics[width=150mm]{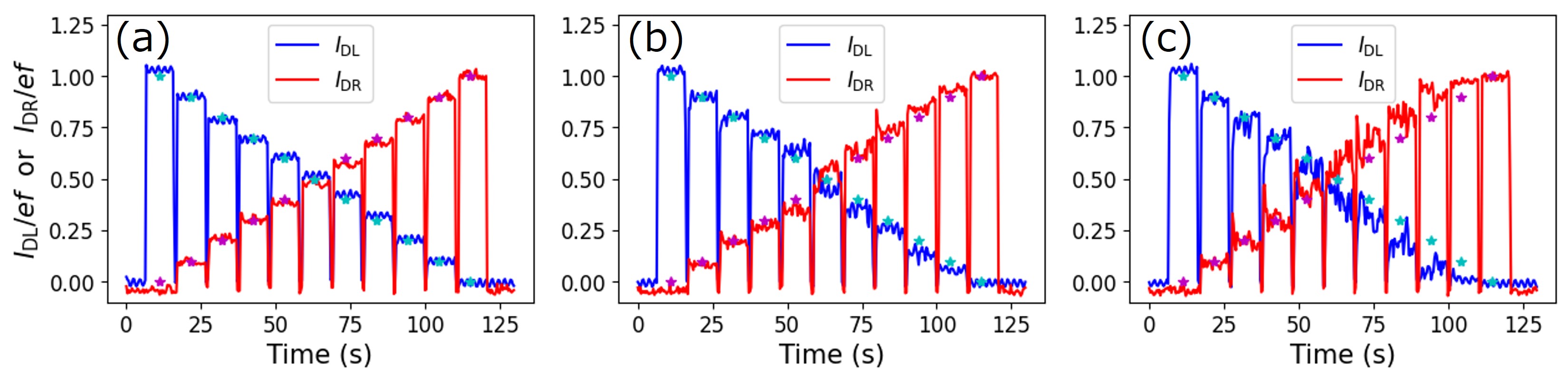}
\caption{\label{0010} Results of electron routing at (a) 1 MHz, (b) 10 MHz, and (c) 100 MHz. The blue and red plots show $I_{\rm{DL}}$ and $I_{\rm{DR}}$, respectively. The cyan and magenta stars ($\star$) show the programmed answer for the ratio of $I_{\rm{DL}}$ and $I_{\rm{DR}}$. {The dips of currents are due to the one second interval inserted when switching single-electron routing ratios.} {The voltage conditions are the same as that of Fig.~\ref{0005}. The phase conditions are $\phi_{\rm{Pump}}=\phi_{\rm{L}}=\phi_{\rm{R}}+\pi$, and $\phi_{\rm{Assist}}$ is varied.}}
\end{figure*}

Finally, the high-speed operation of the single-electron router was verified by shortening the duration of one symbol. Since achieving a current measurement time short enough to trace the routing of every single electron is difficult, we adjusted the routing probability of the path-selection pseudo-random number. We measured the ratio of $I_{\rm{L}}$ and $I_{\rm{R}}$ for 10 seconds per symbol. As shown in Fig.~\ref{0010}, 11 patterns of pseudorandom number sequences with $(R:L) = (0:10), (1:9), (2:8),..., (10:0)$ were generated by varying {the parameter $r$, which determine the routing ratio} from 0 to 1 (see Appendix~\ref{AppC}). Fig.~\ref{0010}(a-c) show the results of routing at 1, 10, and 100 MHz, i.e., changing the routing after a 100, 10, and 1 electron period, respectively. They show that the single-electron router worked well at all routing frequencies. Moreover, Fig.~\ref{0010}(c) indicates that each electron was routed individually, as the operating frequencies of the SEP and the routing were the same, 100 MHz. The noise in Fig.~\ref{0010}(c) is slightly larger than in the other figures; the reason is considered to be the high-frequency component of the incident rf signal during the phase change. This issue needs to be resolved to achieve higher accuracy in the future.

\section{Summary}
This study described a single-electron router in a QD array that works with a SEP operating at 100 MHz. It also revealed the importance of the role of the assist gate in front of the branching paths. {The assist gate is required to avoid electron stagnation between the SEP and the branching paths. This has similarities with the recently proposed transport strategy for electrons as a conveyor-mode shuttling~\cite{seidler2022conveyor,Langrock2023}, where the electron is transported by the multiple sinusoidal signals with phase change, even though the device structure and driving speed differ.}

Possible applications of the single-electron router include the following. 
(i) Error monitoring for SEP ~\cite{yamahata2011accuracy,Yamahata2014,Wulf2013,Fricke2014,Tanttu2015,ghee2023fidelity, keller1996accuracy,Camarota_2012}; when using a SEP as a current standard, a controllable electron branching path is helpful for determining whether the current is produced with the desired accuracy. It allows the current value to be monitored by performing spot checks at a certain time interval, thereby maintaining a stable current standard. {For such an application, a high-precision measurement setup evaluating the metrological fidelity of a single-electron router is indispensable.}
(ii) Transporting electrons in a spin-based quantum computer~\cite{Flentje2017,Ansaloni2020,gilbert2020single,duan2020remote,Mortemousque2021,borsoi2022shared}; electron routing can be used for electron transport during the preparation and readout steps of a quantum computer in which electron spins are used as qubits, enabling electron transport to a desired QD on a two-dimensional QD array. {We believe that the electron transport technologies, including our single-electron router, are essential for quantum computers, although still in the early research stage.}
(iii) Single-electron splitters or divider~\cite{Bocquillon2012,Bocquillon2013,Dubois2013,Waldie2015,Fletcher2019,Edlbauer2022}; electron routing can be used in electron control elements that split electrons into the desired ratios. Note that whereas an optical beam splitter creates a superposition of photons, a single-electron router separates electrons deterministically. Thus, in contrast to coherent electron beam splitters in electron quantum optics, which are based on the wave nature of single electrons~\cite{bauerle2018coherent,Edlbauer2022,yamamoto2012electrical,takada2019sound,Ito2021}, a single-electron router separates electrons incoherently.

In conclusion, we conducted a proof-of-principle experiment on a single-electron router technology that selects a single-electron path by using a tunable barrier SEP in a QD array. We established the theory behind an evaluation of the routing performance and demonstrated the programmable routing of electrons pumped by the SEP by using an assist gate placed just before the electron branching paths. Finally, we investigated high-speed routing control and found that the router could route individual electrons at 100 MHz.

\begin{acknowledgments}
This work was supported by a JST Moonshot R\&D grant, no. JPMJMS2065.
\end{acknowledgments}

\appendix

\section{\label{AppA}Frequency dependence of SEP}
\begin{figure}[ht]
\includegraphics[width=80mm]{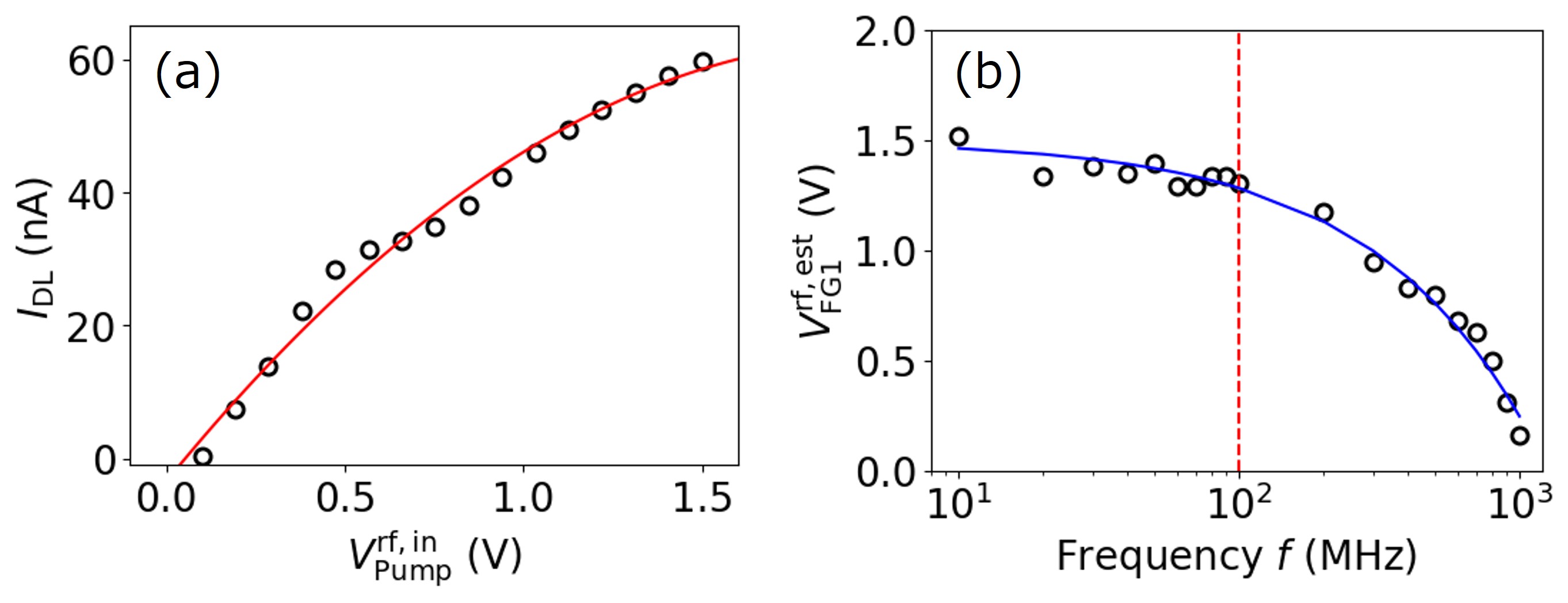}
\caption{\label{A002}Frequency dependence of the pump gate (FG1). (a) Relation between input rf amplitude $V_{\rm{Pump}}^{\rm{rf, in}}$ and the left drain current $I_{\rm{DL}}$ when the bias voltage between the source and the left drain was set to 50 mV, the right drain was opened, and the other gates were set to turn-on by applying 2 V. We also set $V_{\rm{Pump}}^{\rm{dc}}$ = 1.5 V near the $V_{\rm{th}}$ of the pump gate (FG1) and $f$ = 1 MHz. The red curve shows the best-fit result for $I_{\rm{DL}}$ with a quadratic function. (b) Frequency dependence of the estimated rf amplitude $V_{\rm{Pump}}^{\rm{rf,est}}$ calculated using the best-fit curve in (a) when the input amplitude was set to $V_{\rm{Pump}}^{\rm{rf, in}}$ = 1.5 V. The blue solid curve is the best-fit result for $V_{\rm{Pump}}^{\rm{rf, est}}$ with a function $V(f)=1.5-af^b$, where $a=0.0066$ and $b=0.76$ are fitting parameters and $f_{\rm{c}}=252$ MHz is the cutoff frequency defined by $V(f_{\rm{c}})=1.5/\sqrt{2}$. We used 100 MHz (red dashed line) for the SEP operation in this study.}
\end{figure}

\begin{figure}[h]
\includegraphics[width=80mm]{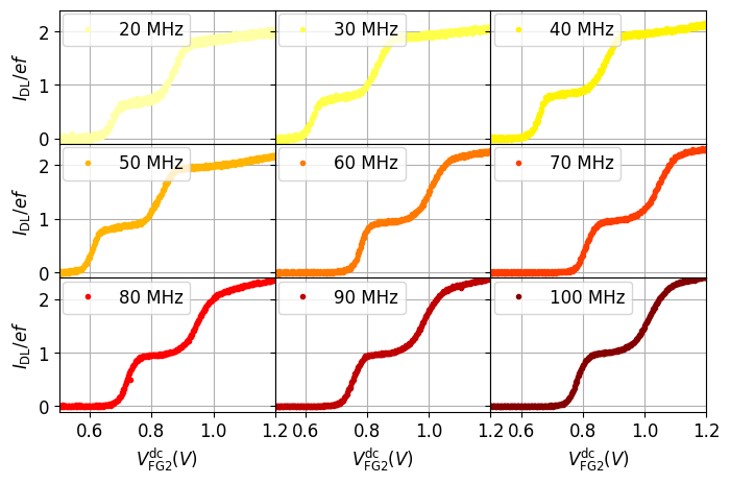}
\caption{\label{A001}Frequency dependence of the SEP plateau. The voltage conditions were optimized for the operation frequency of 100 MHz, as shown in Fig.~\ref{0002}(b).}
\end{figure}

To determine the operating frequency of the SEP, we evaluated the rf transmission characteristics of the pump gate (FG1) as in Ref.~\cite{Utsugi_2023}. From the rf amplitude dependence of $I_{\rm{DL}}$ shown in Fig.~\ref{A002}(a), we obtained the cutoff frequency of the transmission to the pump gate (250 MHz) and determined that a 100-MHz signal could be applied [Fig.~\ref{A002}(b)]. Next, the characteristics of the single-electron plateau were plotted as a function of the driving frequency of the SEP (Fig.~\ref{A001}). The plateau characteristics were plotted by varying only the driving frequency after setting the optimal voltage conditions at a driving frequency of 100 MHz. The current values were normalized by each driving frequency. Clear plateau characteristics appeared at the positions of the single-electron current $I_{\rm{DL}}=ef$ corresponding to each frequency up to about 60 MHz. Below 30 MHz, the plateau characteristics deteriorated, and the deviation of the pumped current values from $ef$ became more noticeable. The reason for this is considered to be the change in the rf amplitude due to the frequency dependence of the transmission loss, as shown in Fig.~\ref{A002}(b).

\begin{figure}[h]
\includegraphics[width=50mm]{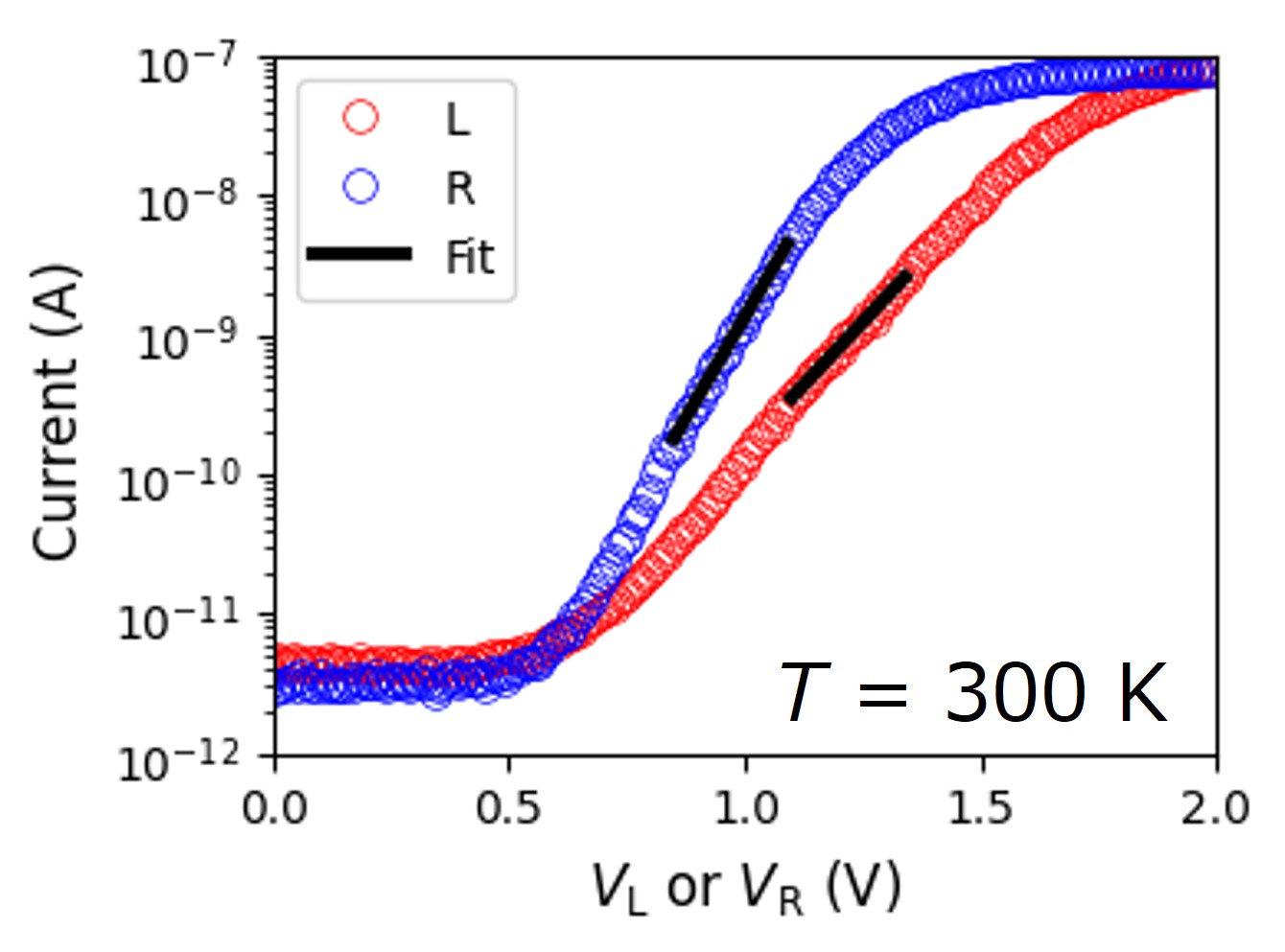}
\caption{\label{A000}{I-V curve of L gate (TG1, red) and R gate (TG2, blue) at room temperature (300 K), where the bias voltage between the right and left drain was set to 50 mV, and the source was set to open. For L (R) gate measurement, the R (L) gate and SGS were set to turn-on by applying 2 V and the other gates were set to 0 V. From the fit lines (black), which are $\exp(aV_{\rm{FG2}}/k_{\rm{B}}T+b)$, where $a$, $b$, $T$ are the lever arm, constant, and temperature, respectively, we obtained $a=$ 0.21 (0.35) eV/V for L (R) gate.}}
\end{figure}

{\section{\label{AppP}Estimation of the lever arm of the switching gates}
We estimated the lever arms of L gate and R gate from I-V curve in the subthreshold regime at 300 K~ \cite{yamahata2019}.
From the parameters extracted from the fits in Fig.~\ref{A000}, the lever arms are estimated as 0.21 (0.35) eV/V for L (R) gate. }

\section{\label{AppD}Validity of the approximation on the final line in Eq.~(\ref{eq4})}
\begin{figure}[h]
\includegraphics[width=80mm]{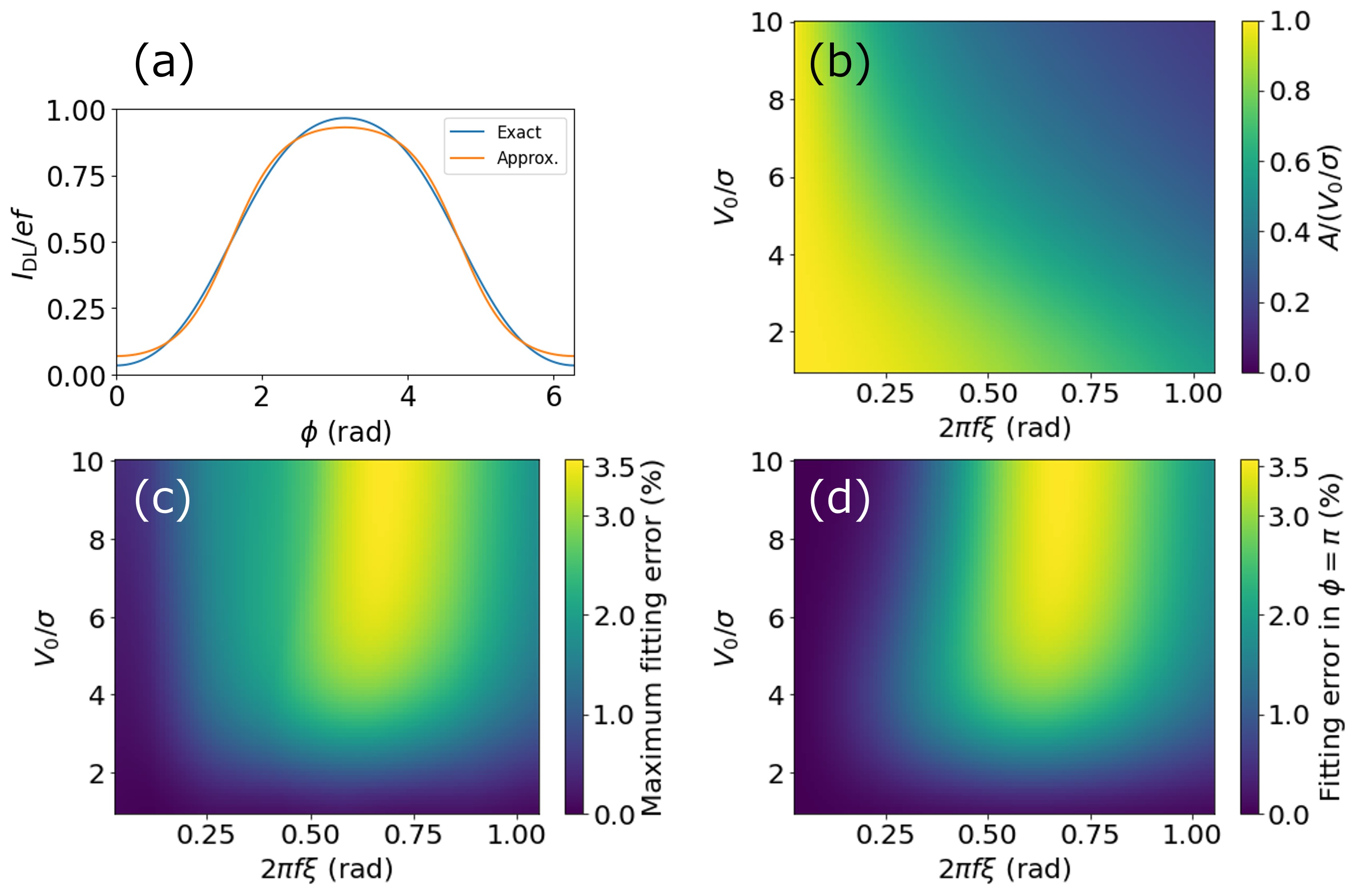}
\caption{\label{A004} (a) Example of the calculation result, where $V_0/\sigma=8$, $\xi=0.7$, and $A=2.6$. The blue line is the calculation result, and the orange line is the fitting curve. (b) Fitting parameter $|A|$ divided by $V_0/\sigma$ plotted as a function of $V_0/\sigma$ and $\xi$. (c) Maximum fitting error in the same calculation with (b). (d) Fitting error in $\phi=0$ in the same calculation as (b) and (c).}
\end{figure}

Here, we numerically show the validity of the relationship described as
\begin{align}
\frac{I_{\rm{DL}}}{ef}
&=\frac{1}{\sqrt{2\pi} \xi} \int^{\tau}_{0} \frac{\exp \left[-\frac{\left(t-\frac{\tau \phi}{2\pi}\right)^2}{2\xi^2}\right]}{\exp \left[-\frac{V_0}{\sigma} \cos(2\pi ft - \phi_{\rm{L}}) \right]+1} dt \notag \\
&\approx \frac{1}{\exp \left[A \cos(\phi-\phi_{\rm{L}})\right]+1}.
\label{eqs1}
\end{align}
We calculated the first line in Eq.~(\ref{eqs1}) as a function of $\phi$ and fit the result with the second line in Eq.~(\ref{eqs1}), where $\phi_{\rm{L}}=0$. In this calculation, $V_0/\sigma$ and $\xi$ were varied within experimentally reasonable ranges. Fig.~\ref{A004}(a) shows an example of the results, where $V_0/\sigma=8$, $\xi=0.7$, and $A=2.6$. The blue line is the calculation result (labeled by `Exact'), and the orange line is the fitting curve (labeled by `Approx.'). Fig.~\ref{A004}(b) shows the results for the fitting parameter $|A|$ divided by $V_0/\sigma$ plotted as a function of $V_0/\sigma$ and $\xi$. This result shows that$|A| \approx V_0/\sigma$ holds for $\xi\rightarrow0$, as the main text describes. $|A|$ decreases by broadening the jitter $\xi$. Thus, the reduction in $|A|$ reflects broadening of the jitter in the electron transport.
Fig.~\ref{A004}(c) shows the maximum fitting error for the same calculation as in Fig.~\ref{A004}(b). The fitting error is evaluated by the absolute value of the difference between the calculated curve and the fitting curve. Since the maximum fitting error is less than 5\%, where $I_{\rm{DL}}/(ef)$=1=100\%, the approximation using Eq.~(\ref{eqs1}) is valid. Furthermore, we investigated the effect of the fitting error on the minimum error rate of the single-electron router $P_{\rm{error}}^{\rm{Router}}$. Fig.~\ref{A004}(d) shows the fitting error (`Exact' - `Approx.') in $\phi=\pi$, which represents $P_{\rm{error}}^{\rm{Router}}$. Since the fitting error is positive over the entire map, we find that `Exact '$\ge$ `Approx.'. The minimum error obtained from `Approx.' overestimates the minimum error obtained from `Exact'; i.e., the actual minimum error is always lower than the evaluation result.

{
\section{\label{AppX}Effective range of parameter $A$}
\begin{figure}[]
\includegraphics[width=60mm]{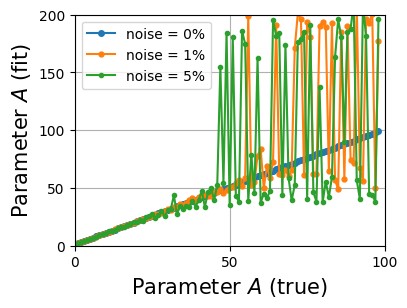}
\caption{\label{A00X}Simulation result for estimation of parameter $A$ by fitting from the curve shown in the approximation in Eq.~(\ref{eq4}), which is sampled at 24 points per 2$\pi$ and added noise as shown in the legend.}
\end{figure}
In our experiments, phases were modulated at 24 steps per 2$\pi$. This step number determines the upper limit of the physically meaningful parameter $A$. The  simulation result shown in Fig.~\ref{A00X} indicates that the estimation of $A$ by fitting is only valid for approximately less than 50 due to the effect of the finite sampling points. For $A > 50$, the fitting accuracy of $A$ is significantly reduced by the experimentally possible noise of a few percent.
}

\section{\label{AppC}Pseudo-random number generation}
The pseudo-random number we used to demonstrate the single-electron router, denoted by $y_n$, was generated with the following recurrence formula: $y_n=\mathrm{Integer}[x_n/(m\times r)]$, where $x_{n+1}=\mod(a x_n, m)$. The values (0,1) taken by $y_n$ correspond to the routing paths to the right and left drains (R, L), respectively. In Fig.~\ref{0009}, each parameter was set to $x_0=4, a=89, m=1993, r=0.5$, and 100 symbols were generated. In addition, as shown in Fig.~\ref{0010}, the parameter $r$ was varied from 0 to 1 in 11 steps.

\bibliography{Router_paper}

\end{document}